\documentclass[manuscript=fullresearchpaper,american,utf8]{beilstein}

\usepackage{xspace}
\nolinenumbers

\begin{document}

\title{Design aspects of Bi$_2$Sr$_2$CaCu$_2$O$_{8+\delta}$ THz sources: optimization of thermal and radiative properties}
\author{Mikhail M. Krasnov}{kmm@kiam.ru}
\affiliation{Keldysh Institute of Applied Mathematics of RAS,
Moscow, Russia}

\affiliation{Moscow Institute of Physics and Technology, 141700
Dolgoprudny, Russia}

\author[1,2]{Natalia D. Novikova}{natasha.tulina@gmail.com}

\author{Roger Cattaneo}{roger.cattaneo@fysik.su.se}
\affiliation{Department of Physics, Stockholm University, AlbaNova
University Center, SE-10691 Stockholm, Sweden}

\author[3]{Alexey A. Kalenyuk}{kalenyuk77@gmail.com}
\affiliation{Institute of Metal Physics of National Academy of
Sciences of Ukraine, 03142 Kyiv, Ukraine}

\affiliation{Kyiv Academic University, 03142 Kyiv, Ukraine}

\author*[2,3]{Vladimir M.
Krasnov}{vladimir.krasnov@fysik.su.se}

\maketitle

\begin{abstract}
Impedance matching and heat management are important factors
influencing performance of THz sources. In this work we analyze
thermal and radiative properties of such devices based on mesa
structures of a layered high-temperature superconductor
Bi$_2$Sr$_2$CaCu$_2$O$_{8+\delta}$. Two types of devices are
considered, containing either a conventional large single crystal,
or a whisker. We perform numerical simulations for various
geometrical configurations and parameters and make a comparison
with experimental data for the two types of devices. It is
demonstrated that the structure and the geometry of both the
superconductor and the electrodes are playing important roles. In
crystal-based devices an overlap between the crystal and the
electrode leads to appearance of a large parasitic capacitance,
which shunts THz emission and prevents impedance matching with
open space. The overlap is avoided in whisker-based devices. Furthermore, the
whisker and the electrodes form a turnstile (crossed-dipole)
antenna facilitating good impedance matching. This leads to more
than an order of magnitude enhancement of the radiation power
efficiency in whisker-based, compared to crystal-based devices.
These results are in good agreement with presented experimental
data.
\end{abstract}


\section{Introduction}

Tunable, monochromatic, continuous-wave (CW), compact and power-efficient terahertz (THz) sources  of electromagnetic waves (EMW) are
required for a broad variety of applications \cite{Tonouchi_2007}.
However, the key obstacle, colloquially known as ``the THz gap" \cite{Tonouchi_2007}, is caused by a low radiation power efficiency (RPE) of THz sources. Despite a remarkable progress achieved by semiconducting quantum cascade lasers (QCL's) \cite{Slivken_2015,Capasso_2015},
their RPE drops
well below a percent level at low THz frequencies
\cite{Wang_2016,Curwen_2019,Walther_2007}. Furthermore, operation
of QCL's is limited by the thermal smearing of quantum levels, which
becomes significant at frequencies $f \lesssim k_B T/h$, where
$k_B$ and $h$ are Boltzmann and Planck constants and $T$ is the
operation temperature. For room temperature, $T=300$ K, this
happens at $f\simeq 6.25$ THz. QCL's have to be cooled down in order
to operate at significantly lower primary frequencies
\cite{Wang_2016,Curwen_2019,Walther_2007}. Although room
temperature operation of QCL's at low frequencies can be achieved
via mixing and down conversion of higher primary frequencies,
this comes at the expense of dramatic reduction of RPE
\cite{Slivken_2015,Capasso_2015,Curwen_2019,Belkin_2013,Rosch_2015}.

The layered high-temperature superconductor
Bi$_2$Sr$_2$CaCu$_2$O$_{8+\delta}$ (Bi-2212) may provide an
alternative possibility for creation of cryogenic CW THz sources
\cite{Ozyuzer_2007,Benseman_2013,Welp_2013,Kashiwagi_2015,HBWang_2015,Borodianskyi_2017,Sun_2018,Kashiwagi_2018,HBWang_2019,Kuwano_2020,Tsujimoto_2020,Saiwai_2020,Delfanazari_2020,Saito_2021}.
Bi-2212 represents a natural stack of atomic scale intrinsic
Josephson junctions (IJJ's)
\cite{Kleiner,Krasnov_1999,Katterwe_2009,Katterwe_2010}. Josephson
junctions have an inherently tunable oscillation frequency,
$f_J=(2e/h)V$, where $e$ is electron charge and $V$ is the bias
voltage per junction. The frequency is limited only by the
superconducting energy gap, which can be in excess of $30$ THz for
Bi-2212 \cite{Krasnov_2000,SecondOrder}. A broad range tunability
of emission in the whole THz range $1-11$ THz has been
demonstrated from small Bi-2212 mesa structures \cite{Borodianskyi_2017}.

Operation of Josephson emitters is limited by two primary
obstacles: self-heating and impedance mismatching. Josephson devices
stop operating when their temperature exceeds the superconducting
critical temperature $T_c$. Self-heating in Bi-2212 mesa
structures has been intensively studied
\cite{Krasnov_2001,Krasnov_2002,Heating_2005,SecondOrder,Yurgens_2011,Kakeya_2012,Kakeya_2014,Asai_2014,Rudau_2015,Benseman_2015,Rudau_2016,Oikawa_2020}.
Although $T_c$ of Bi-2212 may be quite high, up to $\simeq 95$ K
\cite{SecondOrder}, self-heating is substantial due to a low heat conductance of superconductors. Self-heating limits the
maximum bias voltage that can be reached without critical
overheating of a mesa and, therefore, the maximum achievable
frequency and emission power. Furthermore, as pointed out in Ref.
\cite{Cattaneo_2021}, self-heating creates a general limitation
for the maximal achievable emission power for any cryogenic device
(not only superconducting). Taking into account the limited
cooling power of compact cryo-refrigerators (sub-Watt at low $T$),
a device with RPE $\sim 1\%$ would not be able to emit
significantly more than 1 mW. Therefore, larger emission power from
cryogenic sources may only be achieved via enhancement of RPE. The
maximum achievable RPE is $50\%$ in the case of perfect matching
of the device microwave impedance with that for open space
\cite{Krasnov_2010}. However, the reported RPE of Bi-2212 THz sources is much
smaller \cite{Borodianskyi_2017} due to a significant impedance mismatch. Therefore, improvement of THz sources requires proper design of cooling
elements to handle self-heating, and impedance matching microwave antennas,
to improve RPE.

In this work we analyze design aspects of THz sources based on
Bi-2212 mesa structures. Thermal and radiative properties are
studied for two types of devices containing either a conventional
large single crystal, or a needle-like whisker. We present numerical
simulations for various geometrical configurations and parameters
and make a comparison with experimental data. It is demonstrated
that the structure and the geometry of both the superconductor and
the electrodes are playing important roles. Electrodes provide an
effective heat sink channel and help in reduction of self-heating.
They also influence radiative properties. However, this influence is opposite
for crystal-based (worsen) and whisker-based (improve)
devices. The superconductor geometry is also crucial. Devices
based on large crystals suffer from a large parasitic capacitance
at the overlap between the crystal and the electrodes. It prevents
good impedance matching and reduces RPE. The overlap is avoided in
whisker-based devices. Moreover, the whisker itself, together with
electrodes, forms a turnstile (crossed-dipole) antenna,
facilitating good impedance matching. We show that this can lead to more that an
order of magnitude enhancement of RPE, compared to crystal-based
devices. Those results are in good agreement with experimental
data, which demonstrate that THz emission from whisker-based
device is much larger than from crystal-based devices with the
same geometry.

\section{Experimental results}

\begin{figure*}
\caption{Optical images of (a) whisker and (b) crystal-based
devices with similar electrode geometries. (c) A sketch of both
devices. } \label{fig:Samples}
\includegraphics[width=15cm,keepaspectratio]{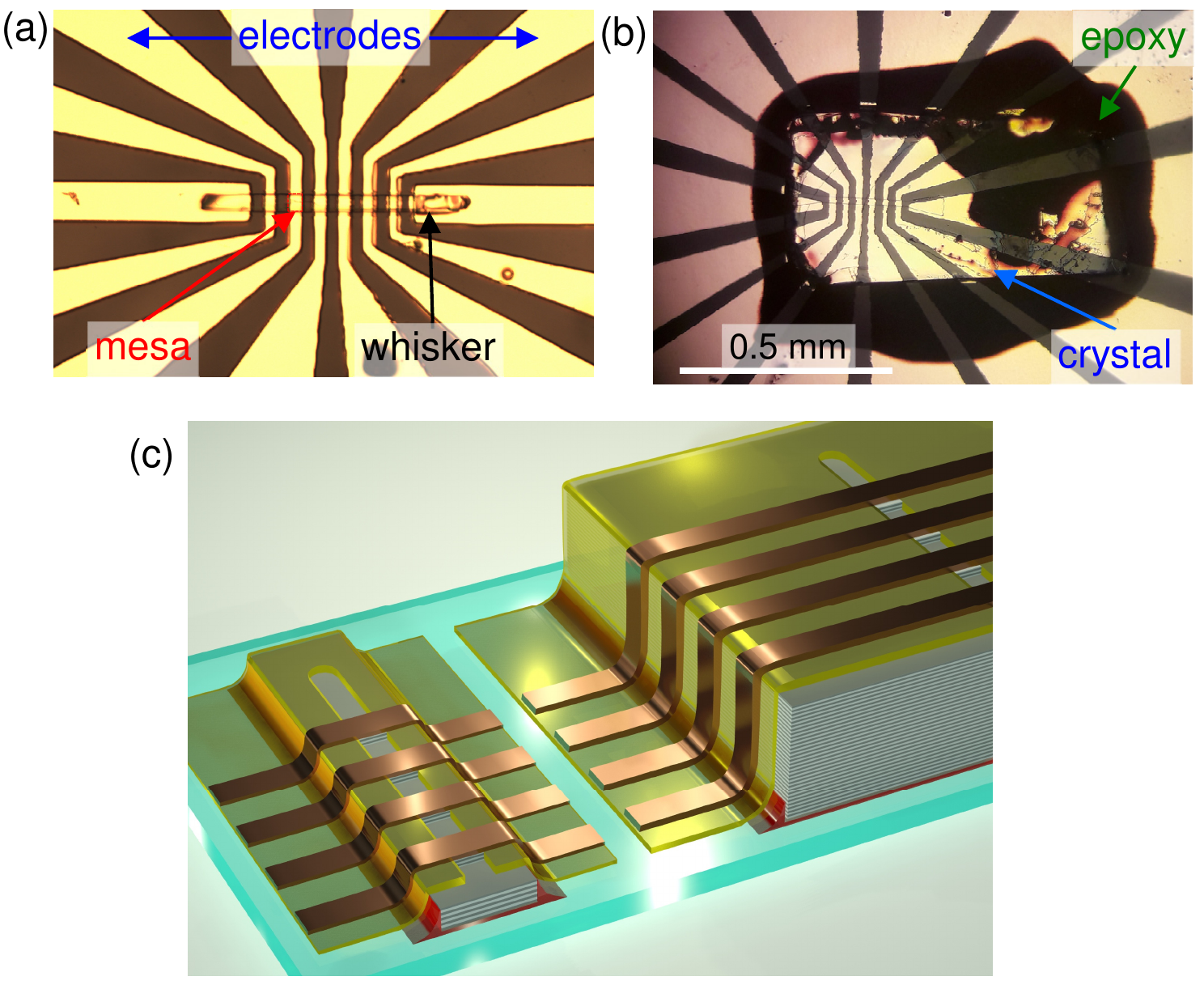}
\end{figure*}

Figures \ref{fig:Samples} (a) and (b) show optical images of two studied
devices. They have a similar geometry and were fabricated using
the same procedure. The main difference is that the device in (a)
is made using a whisker; and in (b) using a conventional single
crystal. Panel (c) shows sketches of both devices.
Bi-2212 whiskers have typical aspect ratios 100:10:1 in $a$, $b$,
and $c$ crystallographic directions \cite{Matsubara_1989}. Our
whiskers have typical dimensions of several hundreds of microns in
$a$, $20-40~\mu$m in $b$ and just few $\mu$m in the $c$-axis
direction. The conventional single crystal is much larger with sizes of almost a mm$^2$ in the $a-b$ plane and several hundreds of micrometers in the $c$-direction.

The fabrication process starts by gluing a corresponding crystal
on a $5\times 5$ mm$^2$ sapphire substrate using an epoxy glue.
The crystal is cleaved at ambient conditions. After that the
sample is immediately put into a deposition chamber and a
protective gold layer $\sim 60-80$ nm is deposited to avoid
surface passivation. Next, a line pattern in photoresist is made
with the length $100-200~\mu$m and the width $5-15~\mu$m on a flat
portion of Bi-2212 surface, followed by Argon ion etching of
unprotected parts of Au and Bi-2212, deposition of insulating
SiO$_2$ or CaF$_2$ layers and a lift-off of the photoresist at the
line. The depth of Bi-2212 etching at this stage ($d_m \sim 200-400$
nm) defines the height of mesas and the number of IJJ's in the
device, $N=d_m/s$, where $s\simeq 1.5$ nm is the interlayer spacing
between double CuO layers in Bi-2212. After that top metallization
Ti/Au layer with the total thickness $\sim 200$ nm is deposited.
Finally several electrodes, crossing the line in a perpendicular
direction, are made by photolithography and Ar-ion etching. Mesa
structures are formed at the overlap between the line and the
electrodes, as indicated in Fig. \ref{fig:Samples} (a).

\begin{figure*}
\caption{ Current-Voltage characteristics of mesa
structures on (a) whisker and (b) crystal-based devices. (c) and (d) represent on-chip generation-detection experiment for (c) whisker and (d) crystal-based
devices. Here an ac-resistance of the detector mesa is shown as a
function of the total dc-dissipation power, $P_{gen}$, of the
generator mesa. For the whisker-based device (c) a profound emission occurs at the step in the $I$-$V$, marked in (a). For the crystal-based device (d) only a small monotonic increment of $R_{det}$ vs. $P_{gen}$ is observed, caused by a gradual self-heating. 
} \label{fig:Data}
\includegraphics[width=15cm,keepaspectratio]{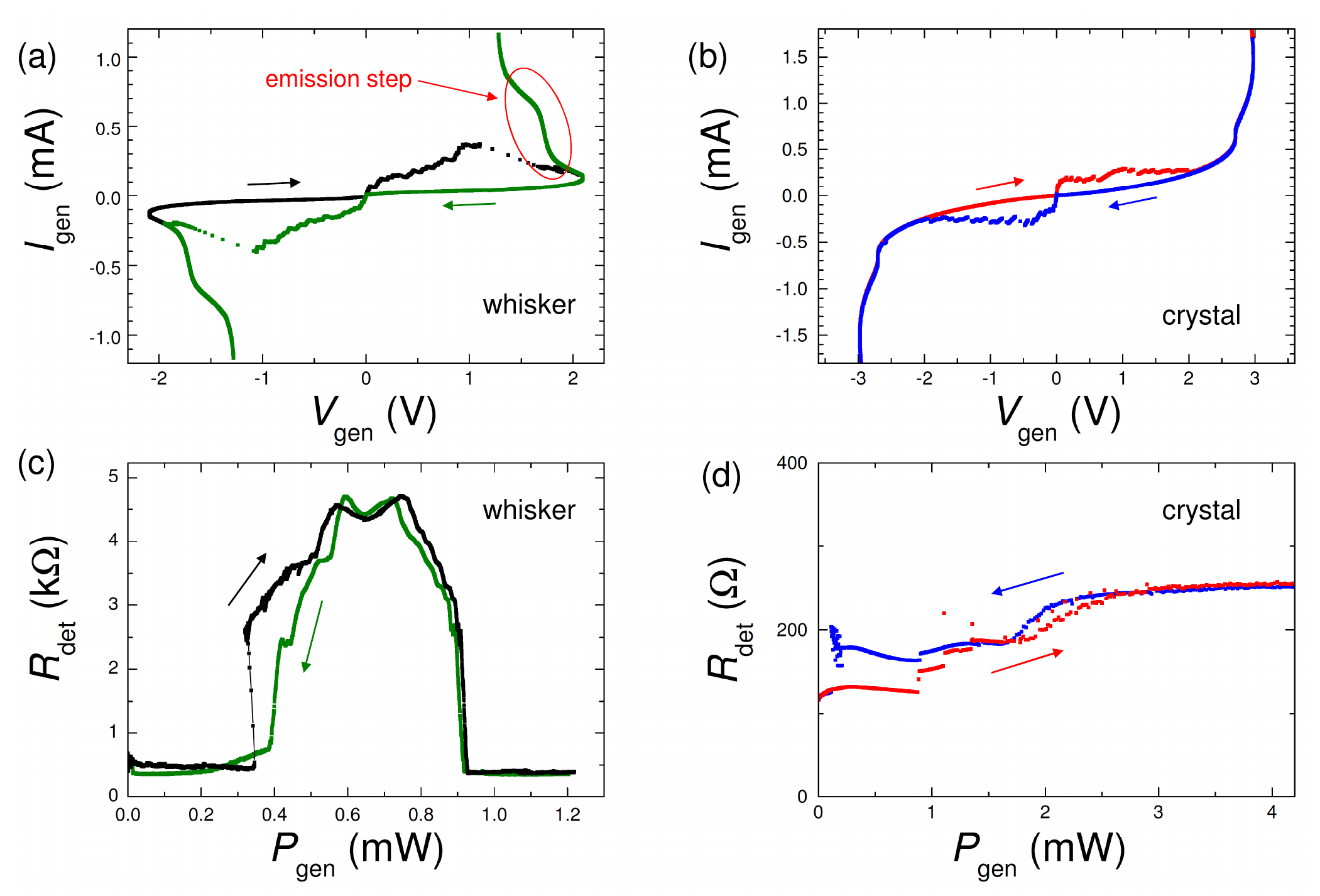}
\end{figure*}

Figures. \ref{fig:Data} (a) and (b) show current-voltage ($I$-$V$)
characteristics of mesas at whisker and crystal-based devices,
respectively. The $I$-$V$'s are fairly similar. They contain
multiple branches due to one-by-one switching of IJJ's from the
superconducting to the resistive state. The are 
$N \sim 200$ and $\sim 300$ IJJ's in whisker and crystal mesas, respectively. Both
the whisker and the crystal have similar suppressed
$T_c\sim 65-70$ K and low critical current densities of IJJ's, $J_c\sim 100$ A/cm$^2$, indicating a strongly underdoped state of Bi-2212 \cite{Jacobs_2016}.

Radiative properties of our whisker-based devices were analyzed in
Ref. \cite{Cattaneo_2021}. A significant EMW emission at $f\simeq
4.2$ THz with a record-high RPE reaching 12$\%$ was reported. The
emission occurs at the step in the $I$-$V$, marked in Fig.
\ref{fig:Data} (a). 
In Fig. \ref{fig:Data} (c) we show results of in-situ THz
generation- detection experiment on a whisker-based device. We follow the procedure, developed in
Ref. \cite{Borodianskyi_2017}, where details of the technique can
be found. We use one mesa with the $I$-$V$ like in Figs.
\ref{fig:Data} (a) as a generator, and another mesa on the same device as a
switching current detector. The detector mesa is biased by a small
ac-current and the generator by a dc-current in the same range as
in Fig. \ref{fig:Data} (a). Fig. \ref{fig:Data} (c) shows
the ac-resistance of the detector mesa, $R_{det}$, as a function
of dissipation power in the generator mesa, $P_{gen}=I_{gen}V_{gen}$. It
is anticipated that self-heating is monotonic (approximately
linear) with the dissipation power, while the emission is
nonmonotonic \cite{Borodianskyi_2017,Cattaneo_2021} because it
occurs at certain bias voltages, corresponding to geometrical
resonances in the mesa
\cite{Borodianskyi_2017,Krasnov_1999,Katterwe_2010,Kashiwagi_2018}. From Fig.
\ref{fig:Data} (c) it is seen that a profound EMW emission occurs in a whisker-based mesa, leading to more than an order of magnitude enhancement of $R_{det}$. The emission occurs in a specific bias range, corresponding to the step in the $I$-$V$, marked in Fig. \ref{fig:Data} (a). To avoid repetitions we address the reader to Ref. \cite{Cattaneo_2021} for more details.  

In Fig. \ref{fig:Data} (d) we show similar generation-detection data for the crystal based device from Figs. \ref{fig:Samples} (b) and \ref{fig:Data} (b). In contrast to the whisker-based device, here we observe only a small monotonic increment of $R_{det}$ 
with increasing $P_{get}$, which is the consequence of self-heating. On top of it there may be a small non-monotonic signal at $0.5$ mW $\lesssim
P_{gen}\lesssim 1.5$ mW, which can be attributed to THz emission.
This is qualitatively similar to results reported earlier for
small mesas on crystal-based devices \cite{Borodianskyi_2017}. For
whisker-based mesas the ratio of emission to self-heating
responses is quantitatively different: The emission peak
$R_{det}(P_{get})$ is much larger than the monotonic self-heating
background. This indicates a much larger RPE in whisker-based devices \cite{Cattaneo_2021}.

\section{Numerical results}

To understand the reported difference between crystal and
whisker-based devices and to suggest possible optimizations of THz sources, we
performed numerical modelling using a 3D finite element software Comsol Multiphysics. Below we present simulations of thermal and radiative properties calculated
using Heat Transfer and RF modules, respectively. Presented simulations contain several simplifications and, therefore, are not aiming to self-consistently predict the extent of self-heating, $\Delta T$, or the radiative power. Their goal is to reveal general trends and geometrical factors, contributing to design aspects of Bi-2212 THz sources.

\subsection{Modelling of heat transfer}

Accurate analysis of self-heating in Bi-2212 mesas is a complex
non-linear problem
\cite{Krasnov_2002,Heating_2005,SecondOrder,Yurgens_2011,Rudau_2015,Rudau_2016}.
Simulations presented below are made for the base temperature
$T_0=10$ K and for sizes similar to actual devices, shown in
Fig. \ref{fig:Samples}: substrate $5\times 5 \times 0.3$
mm$^3$, crystal $1\times 1 \times 0.3$ mm$^3$, whisker $300\times 30 \times 3~\mu$m$^3$ and mesa $30\times 30 \times 0.3~\mu$m$^3$. The thickness of gold electrode 200 nm. The thickness of epoxy layer, $d_e$, depends on the quantity of applied glue, area of the crystal, experience and luck. For whisker devices we managed to reduce it to $d_e \lesssim 1 \mu$m. To do so we use a tiny amount of epoxy and also squeeze it out by pressing the whisker upon gluing. This procedure is effective for whiskers due to their small area. For large crystals, requiring more epoxy, this is more difficult and the remaining epoxy layer is usually thicker. For this reason we assume the epoxy thickness $d_e=1~\mu$m for whisker and $d_e=5~\mu$m for crystal -based devices. 

The monocrystalline sapphire substrate has a very good thermal conductivity, $\kappa$, at low $T$. The substrate is well thermally anchored with the boundary condition at the bottom surface
$T=T_0$. Due to large $\kappa$, temperature
variation in the substrate is negligible. 
Therefore, we use a constant $\kappa=3000$ W K$^{-1}$m$^{-1}$ for the substrate, corresponding to a monocrystalline sapphire at $T\simeq 10$ K \cite{Lytvynov_1970}. To the contrary, the epoxy used for gluing
Bi-2212 crystals, has a poor heat conductance at low $T$. We do
not consider its $T$-dependence because it acts just as a heat
blocking layer, which we assume to have $\kappa_e=0.0025$
W K$^{-1}$m$^{-1}$. 
On the other hand, it is necessary to take into account actual $\kappa(T)$ dependencies for the other two
materials, Bi-2212 and polycrystalline gold electrodes. At low $T$
both have linear $\kappa(T)$. For Bi-2212 we assume
$\kappa_{ab}(T)=0.1~T$(K) W K$^{-1}$m$^{-1}$ \cite{Yurgens_1991} with
an anisotropy $\kappa_{ab}/\kappa_c=8$ \cite{Crommie_1991}. For a
polycrystalline gold thin film we use $\kappa (T)=3~T$(K)
W K$^{-1}$m$^{-1}$ \cite{Yurgens_2011}. The heat is produced in the mesa volume with the total power of 1 mW and uniform density.

\begin{figure*}
\caption{Heat transport in a whisker-based device without
electrodes. (a) A sketch of the device and (b) a cross-section
through the mesa (not in scale). (c-e) Calculated temperature
distribution for the device in vacuum. (f-h) The same for the
device in exchange He gas.} \label{fig:HeatNoEl}
\includegraphics[width=16cm,keepaspectratio]{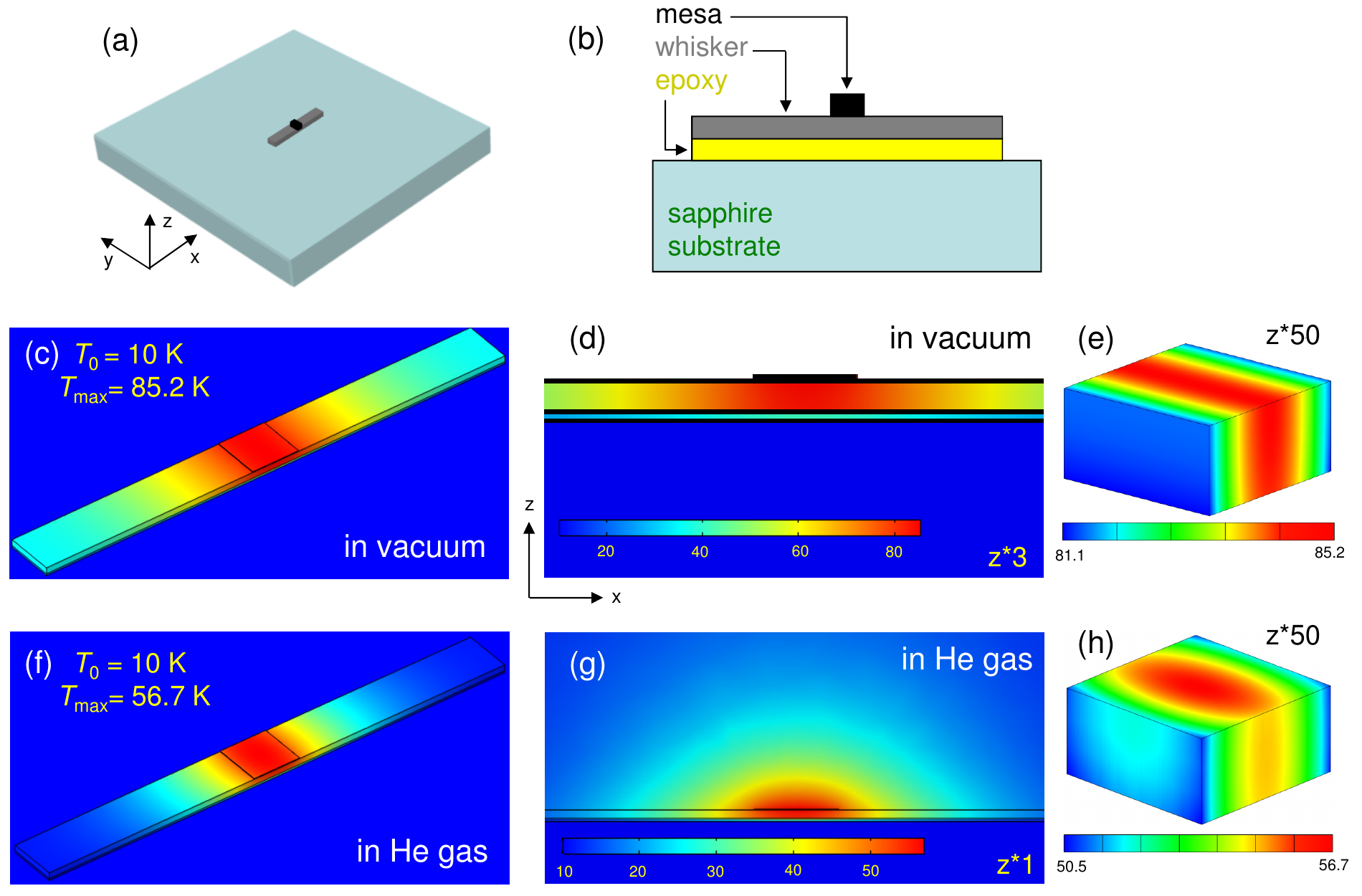}
\end{figure*}

Figure \ref{fig:HeatNoEl} represents heat-transfer simulations for
a whisker without an electrode. Panels (a) and (b) show sketches
of the device and the $x$-$z$ cross-section through the mesa (not
in scale). Figs. (c-e) show the temperature distribution for the
case when the sample is placed in vacuum: (c) top view, (d)
$x$-$z$ cross-section through the mesa (stretched by a factor 3 in
the vertical direction), and (e) in the mesa
(stretched by a factor 50 in the vertical direction). In this case
the heat can only sink into the substrate. As seen from Fig.
\ref{fig:HeatNoEl}(d), the epoxy layer between the substrate and
the whisker blocks heat flow into the substrate and causes
substantial heating of the whole whisker with the maximum
temperature in the center of the mesa reaching $T_{max}=85.2$ K.
Figs. \ref{fig:HeatNoEl}(f-h) show simulations for the same device
in the exchange  $^4$He gas. Clearly, it helps to cool down the
device, although self-heating still remains substantial,
$T_{max}=56.7$ K.

\begin{figure*}
\caption{Heat transport in a whisker-based device with an
electrode. (a) A sketch of the device and (b) a cross-section
through the mesa (not in scale). (c-e) Calculated temperature
distribution for the device in vacuum. (f-h) The same for the
device in exchange He gas.} \label{fig:HeatEl}
\includegraphics[width=16cm,keepaspectratio]{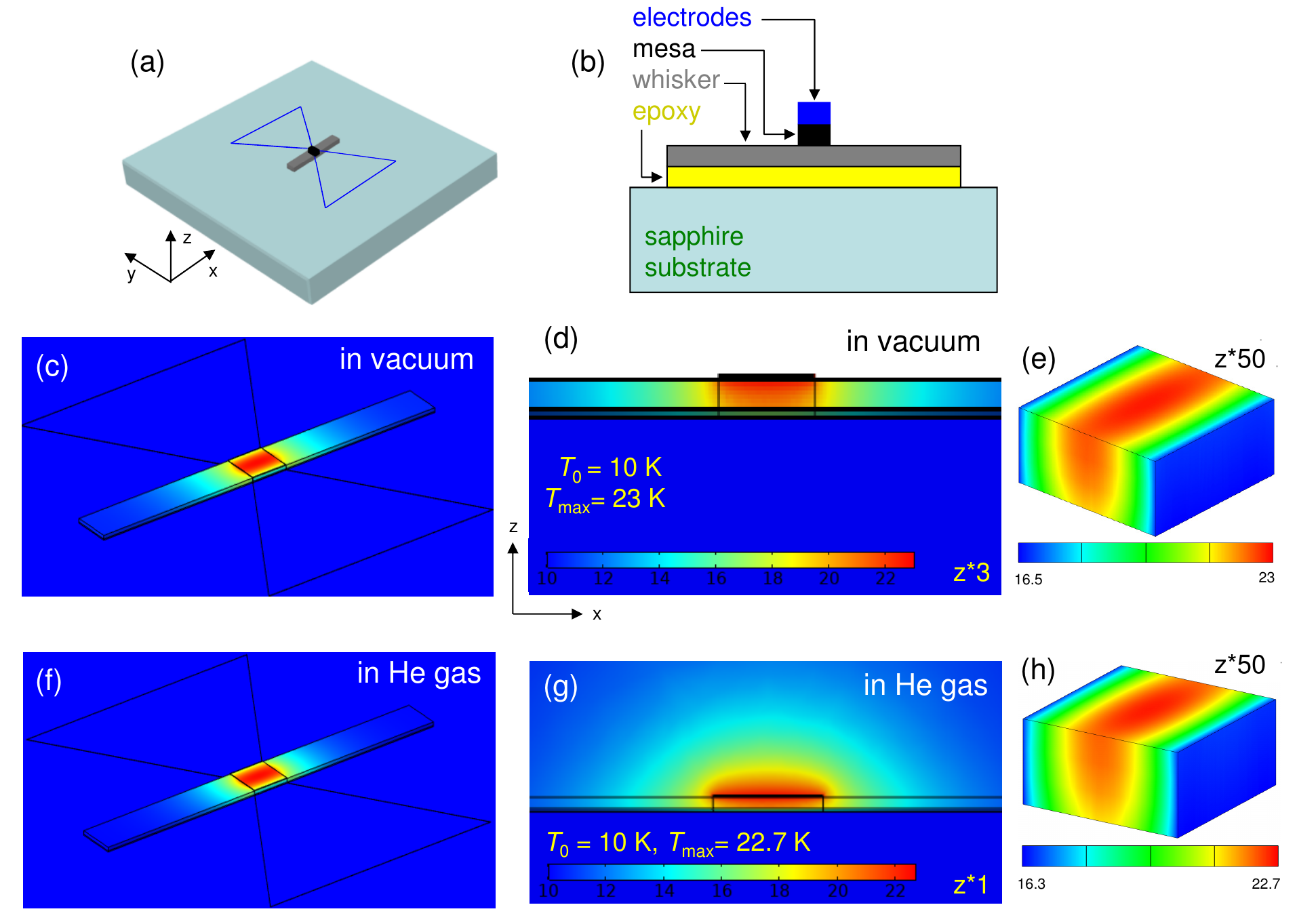}
\end{figure*}

Figure \ref{fig:HeatEl} represents simulations for the
whisker-based device with the top Au electrode. Outside the
whisker the electrode is in a direct contact with the sapphire substrate (no epoxy). This creates a good thermal sink and, as a result, $T_{max}$ falls to $\sim 23$
K. Addition of the exchange gas doesn't play a major role in this
case because the main heat sink channel is provided by the
electrode \cite{Kakeya_2012,Kakeya_2014}, acting as a heat spreading layer \cite{Krasnov_2001}.

\begin{figure*}
\caption{Heat transport in a crystal-based device in vacuum (a)
without electrodes, (b) with electrodes. Left panel represent top
views, middle panels - the $x$-$z$ cross-section through the mesa,
and right panels the mesa (expanded by factor 50 in
$z$-direction).} \label{fig:HeatCrystal}
\includegraphics[width=16cm,keepaspectratio]{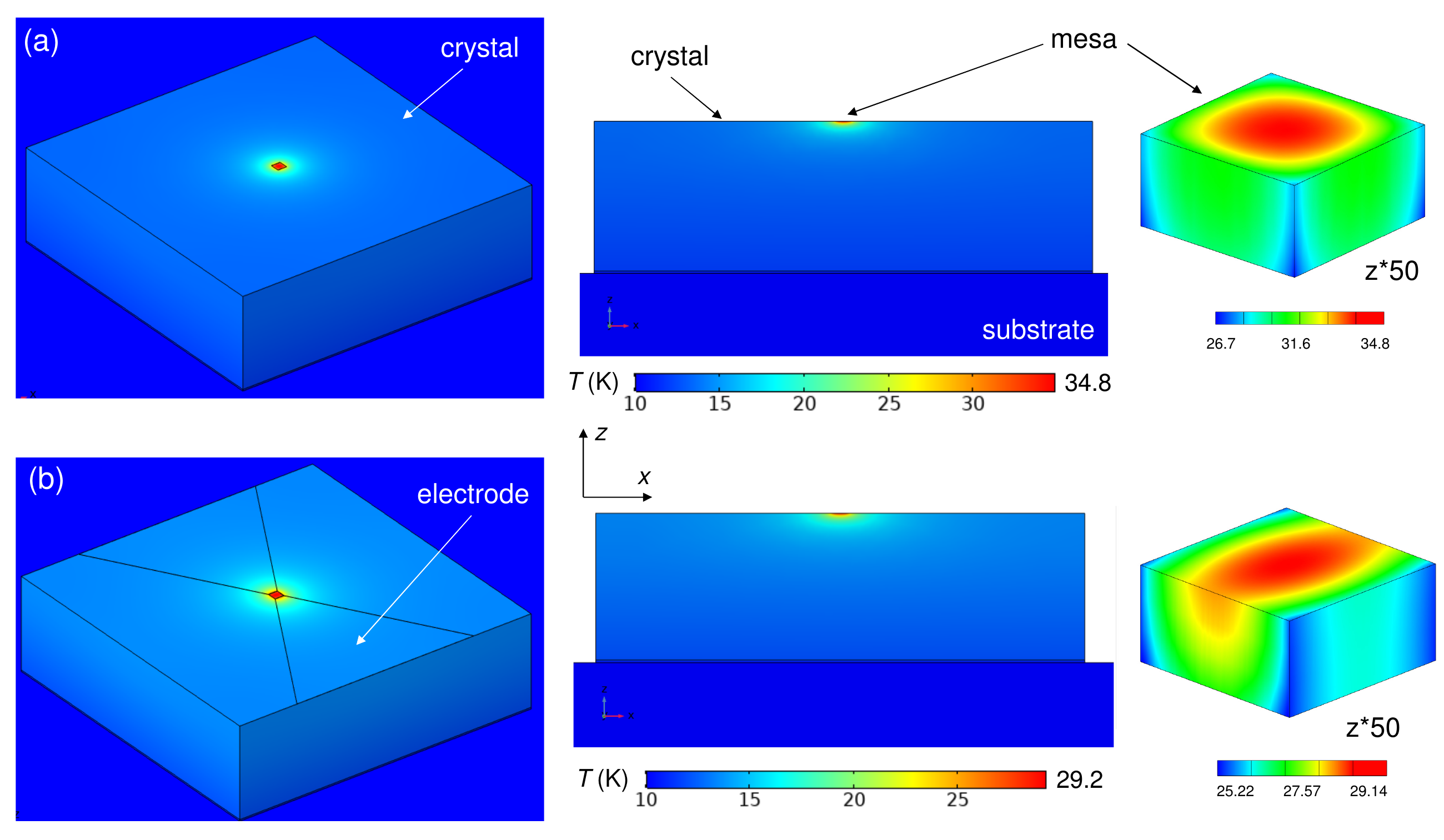}
\end{figure*}

Figure \ref{fig:HeatCrystal} shows temperature distribution in a
crystal-based device in vacuum (a) without electrodes and (b) with
electrodes. The main difference is that unlike in the
whisker-device, Fig. \ref{fig:HeatNoEl}, there is no major
temperature jump in the epoxy layer between the crystal and the
substrate. This occurs because the heat resistance, $R_h = d/(\kappa A)$, is inversely proportional to the area $A$. Due to a much larger crystal area, $R_h$ of epoxy is
negligible despite a low $\kappa$ and larger $d_e=5 ~\mu$m. 
Adding an electrode and He exchange gas further reduces self-heating, but
their effect is not as profound as for the whisker-device, Fig.
\ref{fig:HeatEl}, due to the effective heat sink channel into the
substrate. 

In many cases self-heating is dominated by some bottlenecks. The origin of heat flow hinders is clearly revealed from inspection of thermal gradients in the mesa, where the heat is produced, and by estimation of heat resistances of different elements. For example, from Fig. \ref{fig:HeatNoEl} (e) it is seen that for a bare mesa on a whisker the heat is flowing along the whisker. This occurs because the epoxy layer with a large $R_h=44.4$ K/mW blocks direct (vertical) heat flow into the substrate. However, the maximal self-heating, $\Delta T=75.2$ K, is almost two times larger, implying that there is yet another bottleneck. It is caused by a small $bc$ ($yz$) cross-section area of the whisker. This additional in-plane heat resistance, $R_h\sim 30$ K/mW, corresponds to the effective length of heat spreading along the whisker comparable to the size of the mesa, as can be seen from Fig. \ref{fig:HeatNoEl} (d). For a whisker mesa with an electrode the thermal gradient changes the direction, see from Fig. \ref{fig:HeatEl} (e), indicating that the heat is flowing predominantly along the electrode. For comparison, the $c$-axis heat resistance of the mesa and the whisker are only 1.3 K/mW at $T=20$ K. This implies that a significant reduction of self-heating in whisker devices could be achieved by replacing epoxy with a better heat-conducting material, e.g. by soldering \cite{Benseman_2013}.  
 
For a mesa on a crystal, Fig. \ref{fig:HeatCrystal}, thermal gradient is fairly spherical (taking into account the anisotropy $\kappa_{ab}/\kappa_c=8$). In this case self-heating is dominated by the spreading heat resistance in the crystal \cite{Krasnov_2001,Heating_2005}, $R_h\simeq 1/2L\sqrt{\kappa_{ab}\kappa_c}=23.6$ K/mW at $T=20$ K, where $L=30~\mu$m is the in-plane size of the mesa. For comparison, heat resistance of epoxy is only $2$ K/mW for $d_e=5~\mu$m. Consequently, epoxy is not the major problem for crystal devices (unless it is very thick $d_e>50~\mu$m). For a real device self-heating will depend on the actual geometry, thicknesses and material parameters. However, our analysis indicates that the optimization is much more important and efficient for whisker devices. This is caused by the low intrinsic $c$-axis heat resistance of whiskers due to the small thickness.


\subsection{Modelling of radiative properties}

For calculation of THz properties, a mesa (the source) is modelled
as a lumped port with a fixed voltage amplitude. Unlike
the heat transfer problem, this problem is linear so that the
results directly scale with the source amplitude. To simplify the
perception, we use the amplitude of 1 Volt. Simulations are made
in a sphere with the radius, $R$, which is chosen to be at least
two times larger than the largest device size and the wavelength
in vacuum. A perfectly matching layer with the thickness $0.1~R$
is added outside the sphere to avoid reflections. We checked that
the presented results do not depend on $R$ and, therefore,
properly describe far-field characteristics.

\begin{figure*}
\caption{Simulated radiative properties at $f=1$ THz for (a)
crystal based device, (b) crystal-based device without electrodes,
and (c) whisker-based device. Left panels show sketches of
devices; middle panels - electric field amplitudes in the $x$-$z$
cross-section through the mesa; right panels represent radiation
patterns for the electric field amplitude in the far-field
(outside the simulation sphere). Note a strong field concentration
between the crystal and the electrode in (a).}
\label{Fig_CrCapWh_1THz}
\includegraphics[width=16cm,keepaspectratio]{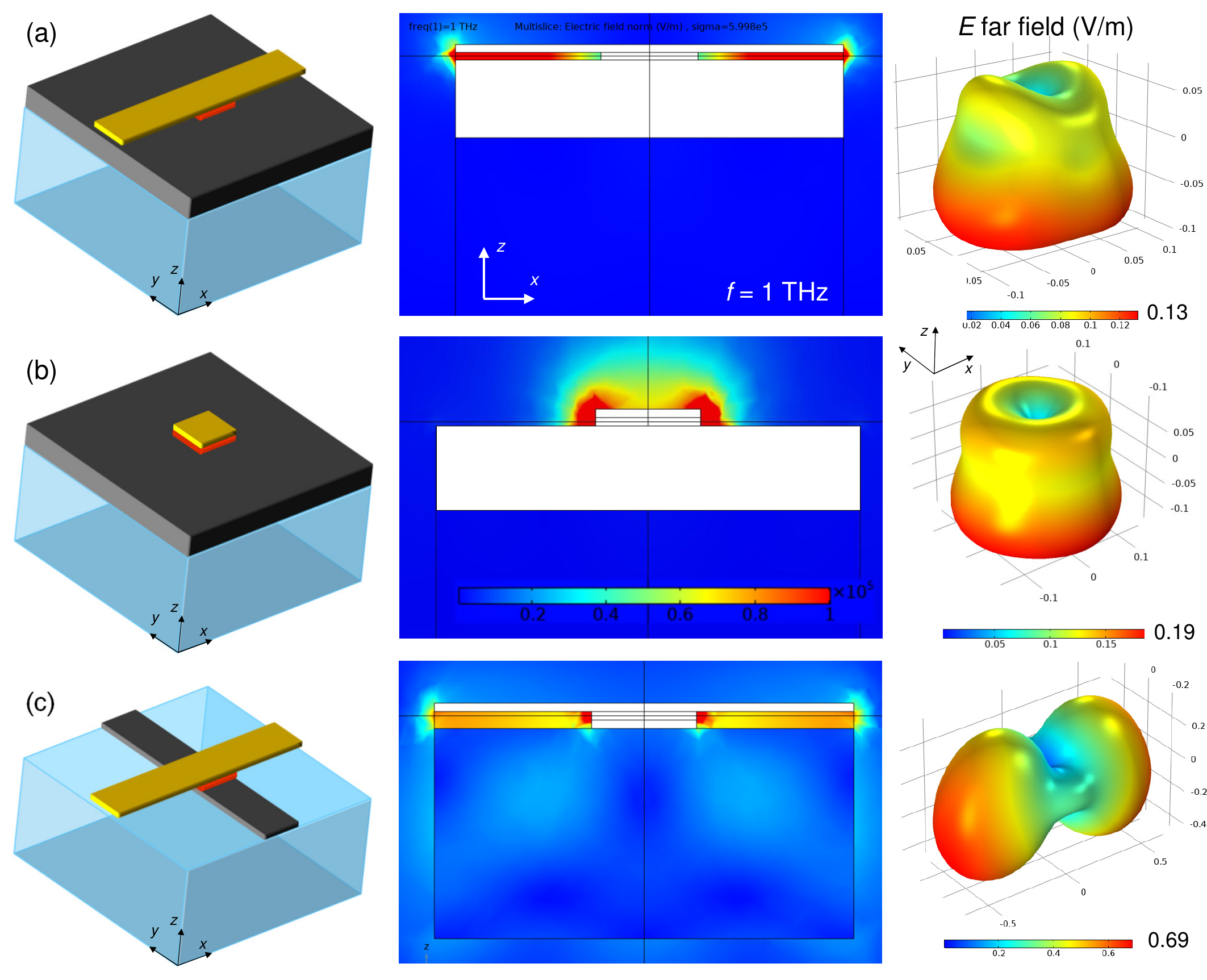}
\end{figure*}

Figure \ref{Fig_CrCapWh_1THz} represents radiative characteristics
for three device geometries, sketched in the leftmost panels: (a)
a mesa (red) on a large crystal (black) with an attached metallic
electrode (yellow), mounted on a dielectric substrate; (b) a mesa
on a large crystal with a capping metallic layer, without
electrode; (c) a mesa on a thin whisker (black) with an attached
electrode. Simulations are performed for $f=1$ THz and the sizes
are selected relative to the wavelength in vacuum,
$\lambda_1=300~\mu$m: the substrate and the in-plane crystal size,
whisker and electrode lengths are $\lambda_1/2=150~\mu$m; the
substrate height is $\lambda_1/4=75~\mu$m; the in-plane mesa size,
whisker and electrode widths are $\lambda_1/8=37.5~\mu$m; the
crystal height is $\lambda_1/10=30~\mu$m; mesa and whisker
heights and the electrode thickness are $\lambda_1/100=3~\mu$m; the
simulation sphere radius $R=2\lambda_1$ and the perfectly matching
layer thickness $0.2~\lambda_1$. The sizes and parameters are
chosen to be similar (but not identical) to studied samples in
order to optimize the mesh size and the calculation time.
Therefore, such simulations serve for a qualitative illustration
of the difference between crystal and whisker-based devices and
the role of the electrodes. 
Electrode and whisker conductivity is
set to $\simeq 6\times 10^5~ (\Omega m)^{-1}$ and relative
dielectric permittivity of the substrate $\epsilon_r=10$.
First we consider the case without dielectric losses, $\tan(\delta)=0$. Middle
panels in Fig. \ref{Fig_CrCapWh_1THz} show local distributions of
electric field amplitudes in the $x-z$ cross-section, going
through the mesa. The same color scale is used, indicated in the
middle panel of Fig. \ref{Fig_CrCapWh_1THz} (b). Rightmost panels
represent far-field radiation patterns (directionality diagrams)
of the electric field amplitude outside the simulation sphere.

From comparison of middle panels in Figs. \ref{Fig_CrCapWh_1THz}
(a) and (c) it can be seen that the electric field distribution is
significantly different. In the crystal-based device the field is
locked between the electrode and the crystal. This occurs because
the electrode is laying on top of the crystal, forming together a
parallel plate capacitor. The field is trapped inside this
capacitor and does not go neither in the substrate, nor open space
in the top hemisphere (with exception of small edge fields). If
we take a realistic specific capacitance $C_{\square}\sim 0.1-1$
fF/$\mu$m$^2$ and electrode area $37.5 \times 150~\mu$m$^2$, we
obtain for $f=1$ THz that the capacitive impedance is very small
$|Z_C|=1/2\pi f C\simeq 0.03-0.3~\Omega$, much smaller that the wave
impedance of the free space, $Z_0=\sqrt{\mu_0/\epsilon_0}\simeq
377~\Omega$. This leads to trapping of EMW in the
electrode/crystal capacitance, which shunts open space and
prevents emission.

To the contrary, for the whisker-based device, Fig.
\ref{Fig_CrCapWh_1THz} (c), the field goes out of the mesa as can
be seen from the brighter overall tone of the pattern in the
middle panel. The EMW propagation is particularly well seen in the
bottom hemisphere due to formation of a standing wave pattern in
the substrate. It is induced by reflections at the
substrate/vacuum interfaces caused by a significant difference in
refractive indices. Emission of EMW is associated with a
cross-like structure of the whisker device, as sketched in the
leftmost panel of  Fig. \ref{Fig_CrCapWh_1THz} (c). It obviates
direct overlap of the whisker and the electrode and prevents
appearance of the large parasitic capacitance. This cross-like
structure resembles the turnstile (crossed-dipole) antenna
geometry, which facilitates good impedance matching with open
space.

The difference between crystal and whisker-based devices is also
reflected in the far-fields characteristics, shown in the
rightmost panels of (a) and (c). The maximum field amplitudes,
$E_{max}$, marked in bottom right corners, are significantly
different: 0.13 V/m for crystal and 0.69 V/m for whisker-based
device. Since the emitted power 
is proportional to $E_{max}^2$, the RPE of the whisker-based
device is almost 30 times larger than for the crystal-based. This
indicates a good impedance matching of the whisker device and a
poor matching for the crystal device. To further demonstrate the
detrimental role of the parasitic electrode/crystal capacitor, in
Fig. \ref{Fig_CrCapWh_1THz} (b) we considered the case with a mesa
on a crystal without electrode and only with the capping top layer
on the mesa. Such configuration is relevant for large mesas,
contacted by a bonding wire \cite{Ozyuzer_2007}. Remarkably, the
far-field emission is larger, $E_{max}=0.19$ V/m, in the absence
of the electrode. This clearly shows that the electrode on top of
the crystal does not help in impedance matching. To the contrary,
it makes things worse due to formation of the large parasitic
capacitance, shunting the EMW.

\begin{figure*}
\caption{Variation of radiative properties with increasing
dielectric losses $\tan(\delta)=$ 0 (top row), 1 (middle row), and
2 (bottom row) for (a) crystal-based (two leftmost columns) and
(b) whisker-based devices (two rightmost columns). Simulations are
made at $f=1$ THz. Note a rapid suppression of the far-field
amplitudes in crystal-based devices.} \label{fig:tg_delta}
\includegraphics[width=16cm,keepaspectratio]{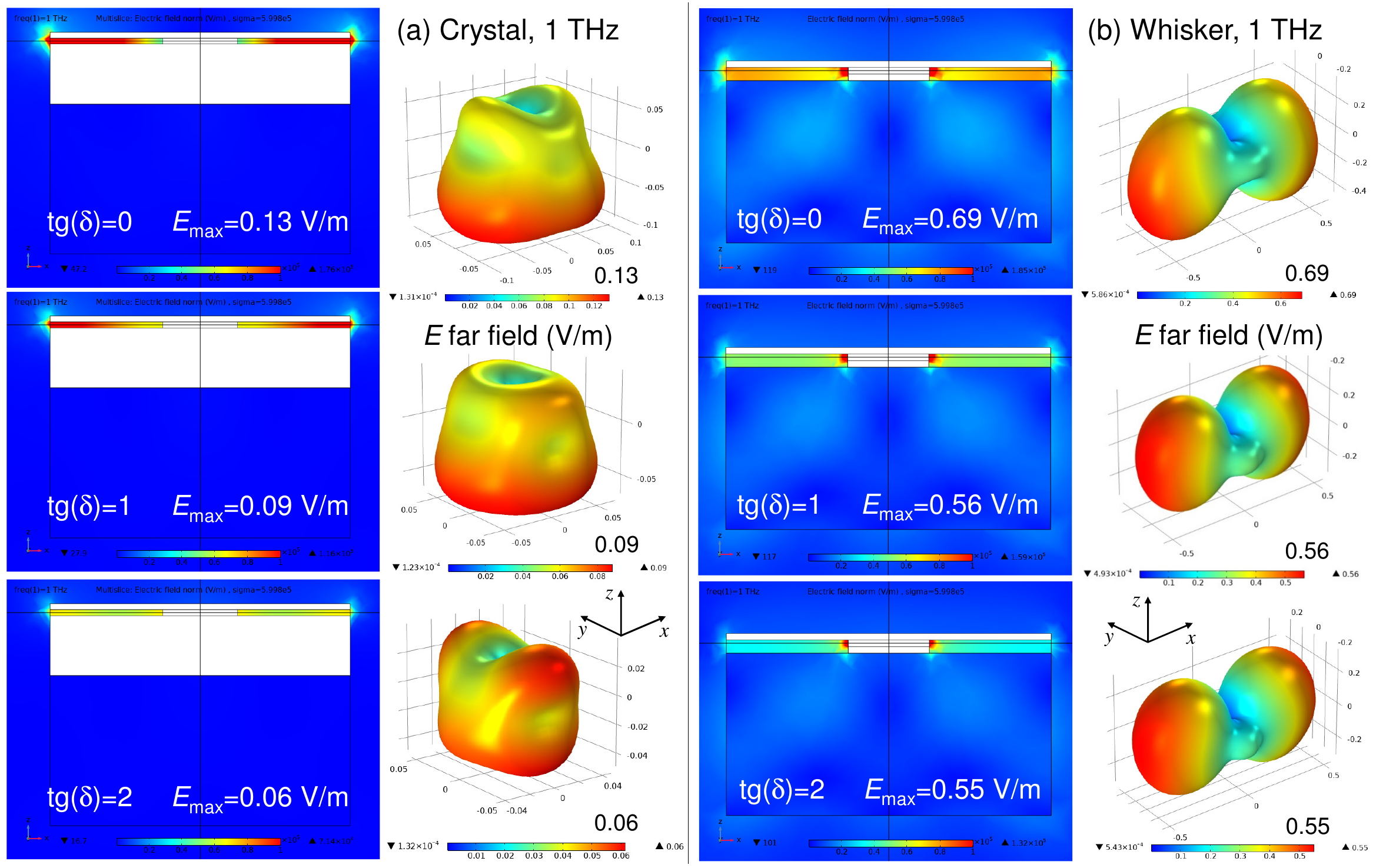}
\end{figure*}

Simulations presented in Fig. \ref{Fig_CrCapWh_1THz} are made for
ideal dielectrics with $\tan(\delta)=0$.
The detrimental role of parasitic crystal/electrode
capacitance becomes much more pronounced if we take into account
dielectric losses, which can be significant at THz frequencies. In
Figure \ref{fig:tg_delta} we show variation of radiative
properties of (a) crystal-based and (b) whisker based devices upon
increasing dielectric losses in the insulating layer between the
crystal and the electrode for crystal-based device and substrate
and electrode for whisker-based device: $\tan(\delta)=0$ (top),
$\tan(\delta)=1$ (middle), and $\tan(\delta)=2$ (bottom row of
panels). It is seen that for whisker-based device dielectric
losses only slightly reduce $E_{max}$ from 0.69 V/m for
$\tan(\delta)=0$ to 0.55 V/m for $\tan(\delta)=2$. For
crystal-based device the relative reduction is significantly
larger, from 0.13 V/m for $\tan(\delta)=0$ to 0.06 V/m for
$\tan(\delta)=2$. As a result, the ratio of RPE for whisker and
crystal devices increases from $\sim 28$ for $\tan(\delta)=0$, to
$\sim 39$ for $\tan(\delta)=1$ and $\sim 84$ for $\tan(\delta)=2$.
This is a direct consequence of electric field concentration in
the parasitic crystal/electrode capacitance of crystal-based
devices.

\section{Discussion}

Josephson oscillators can provide unprecedented tunability in the
whole THz range at a primary frequency \cite{Borodianskyi_2017}.
However, being cryogenic devices, they are susceptible to
self-heating, which limits both the achievable frequency range and
the emission power. As pointed out in Ref. \cite{Cattaneo_2021},
the maximum emission power is limited by the cooling power of the
device and the radiation power efficiency:
\begin{equation}\label{PTHz}
P_{THz} < P_{cooling} \times RPE.
\end{equation}

Enhancement of the effective cooling power requires implementation
of special cooling elements at the device. Despite a significant
progress in this direction
\cite{Benseman_2013,Ji_2014,Asai_2014,Kashiwagi_2015,HBWang_2015,Rudau_2015,Rudau_2016},
it is unlikely that a single emitter would be able to sustain the
dissipation power above few tens of mW. The tolerable dissipation
power can be significantly enhanced by spreading it between
several smaller emitters \cite{Benseman_2013,Tsujimoto_2020}
because smaller mesa structures are less prone to self-heating
\cite{Krasnov_2001,Heating_2005,SecondOrder,Borodianskyi_2017}.
Such a strategy has been successfully proved for arrays of
Josephson junctions \cite{Barbara_1999,Galin_2018,Galin_2020}, for
which coherent emission from up to 9000 synchronized junctions was
reported \cite{Galin_2018}. Yet, the ultimate dissipation power is
limited by the cooling power of the cryostat itself. For compact
cryorefrigerators it is in the range of 100 mW. As follows from
Eq. (\ref{PTHz}), the source with RPE$=1\%$ (which is good for THz
sources) would not be able to emit more than $P_{TH_z}=1$ mW.
Therefore, further enhancement of the emission power requires
enhancement of RPE. This in turn requires proper microwave design
to facilitate impedance matching with open space. The maximum RPE
in case of perfect matching is $50\%$ \cite{Krasnov_2010},
implying that up to 50 mW emitted THz power could be achieved.

Above we considered design aspects of
THz sources, which contribute to obviation of self-heating and
improvement of impedance matching. Several geometries of Bi-2212
devices were analyzed. It is shown that geometries of both the
Bi-2212 crystal and the electrodes are playing important roles.
Their effect, however, depends on the device type. 

For crystal-based devices with large crystals $\sim 1\time 1$
mm$^2$ in the $ab$-plane, see Fig. \ref{fig:Samples} (b), the size
of the crystal is playing opposite roles in device operation. On
the one hand, a large $ab$-plane area helps to spread heat into
the substrate and reduces self-heating of the device, as seen from
Fig. \ref{fig:HeatCrystal}. On the other hand, it leads to a large
overlap area between the crystal and the top electrode. This
creates a large parasitic capacitance that shunts THz emission
and suppresses RPE.

In whisker-based devices the situation is different. Here the
electrode
provides the main heat sink channel, as shown in Fig.
\ref{fig:HeatEl}. In general, our analysis indicated that self-heating optimization is much more important and efficient for whisker devices due to the low intrinsic $c$-axis heat resistance (caused by the small thickness of the whisker).
Furthermore, the cross-like geometry
prevents an overlap between the whisker and the electrode,
thus obviating the parasitic capacitance. Moreover, the long
whisker and the electrode act as two arms of the crossed dipole
(turnstile) antenna, facilitating good impedance matching with
open space. Operation of whisker based devices \cite{Saito_2021,Cattaneo_2021}, and devices based on stand-alone-mesas with similar cross-like electrodes \cite{Kashiwagi_2018} has been demonstrated by several groups.

The role of the substrate is also different.
In crystal-based devices the large superconducting crystal screens
the EMW, so that there is practically no field in the substrate,
see Figs. \ref{Fig_CrCapWh_1THz} (a) and (b). In this case the
substrate does not influence radiative properties. To the
contrary, for whisker-based device a significant fraction of EMW
is going into the substrate due to its larger dielectric constant.
The difference of dielectric constants of the
substrate and vacuum leads to internal reflections and formation
of standing waves in the substrate, see Fig.
\ref{Fig_CrCapWh_1THz} (c). Therefore, the substrate acts as a
dielectric resonator and may strongly affect the radiation pattern
of the device.

Presented numerical simulations provide a qualitative explanation
of the reported difference in radiative properties of whisker and
crystal-based devices, shown in Figs. \ref{fig:Samples} (a) and
(b). They explain why RPE of whisker-based devices is much larger
(by more than an order of magnitude, as follows from Fig.
\ref{fig:tg_delta}). Those conclusions are in agreement with
experimentally reported RPE, which is in the range of $\lesssim 1\%$
for crystal-based \cite{Benseman_2013,Borodianskyi_2017} and up to
$12\%$ for whisker-based \cite{Cattaneo_2021} devices.

\section{Conclusions}

To conclude, intrinsic Josephson junctions in the layered
high-temperature superconductor Bi-2212 can provide an alternative
technology for creation of tunable, CW THz sources. In this work we
analyzed two main phenomena that limit performance of such
devices: self-heating and low RPE caused by impedance mismatching.
We presented numerical simulations of thermal and radiative
properties of Bi-2212 THz sources based on conventional large
single crystals and needle-like whiskers. Simulations are
performed for various geometrical configurations and parameters. A
comparison with experimental data for crystal and whisker-based
devices is made. It is demonstrated that the structure and the
geometry of both the superconductor and the electrodes are playing
important roles. Crystal-based devices suffer
from a large parasitic capacitance due to an overlap between the
crystal and the electrodes. This prevents good impedance matching
and reduces RPE. The overlap is avoided in whisker-based devices.
Moreover, the whisker and the electrodes forms a turnstile
(crossed-dipole) antenna facilitating good impedance matching with
open space. Our simulations demonstrate that this may enhance the
radiation power efficiency in whisker-based devices by more than
an order of magnitude compared to crystal-based devices, which is
consistent with the experimental data.

\begin{acknowledgements}
We are grateful to A. Agostino and M. Truccato for assistance with
whisker preparation and to A. Efimov and K. Shiianov for
assistance in experiment.
\end{acknowledgements}

\begin{funding}
The work was supported by the Russian Science Foundation Grant No.
19-19-00594. The manuscript was written during a sabbatical
semester of V. M. K. at MIPT, supported by the Faculty of Sciences
at SU.
\end{funding}


\bibliography{THzReferences}

\end{document}